\documentclass[a4paper]{article}

\usepackage{icrc2013}

\title{Diffuse TeV Gamma-Ray Emission in the H.E.S.S. Galactic Plane Survey}

\shorttitle{Diffuse Gamma-Ray Emission in the H.E.S.S. Galactic Plane Survey}

\authors{
K. Egberts$^{1}$,
F. Brun$^{2}$,
S. Casanova$^{3,2}$,
W. Hofmann$^{2}$,
M. de Naurois$^{4}$,
O. Reimer$^{1}$,
Q. Weitzel$^{2}$
for the H.E.S.S. Collaboration, and
Y. Fukui$^{5}$
}

\afiliations{
$^1$ Institut f\"ur Astro- und Teilchenphysik, Leopold-Franzens-Universit\"at Innsbruck, A-6020 Innsbruck, Austria  \\
$^2$ Max-Planck-Institut f\"ur Kernphysik, P.O. Box 103980, D 69029 Heidelberg, Germany \\
$^3$ Unit for Space Physics, North-West University, Potchefstroom 2520, South Africa \\
$^4$ Laboratoire Leprince-Ringuet, Ecole Polytechnique, CNRS/IN2P3, F-91128 Palaiseau, France \\
$^5$ Department of Astrophysics, Nagoya University, Chikusa-ku, Nagoya 464-8602, Japan\\

}

\email{kathrin.egberts@uibk.ac.at}

\abstract{Diffuse $\gamma$-ray emission has long been established as the most prominent feature in the GeV sky. Although the imaging atmospheric Cherenkov technique has been successful in revealing a large population of discrete TeV $\gamma$-ray sources, a thorough investigation of diffuse emission at TeV energies is still pending. Data from the Galactic Plane Survey (GPS) obtained by the High Energy Stereoscopic System (H.E.S.S.) have now achieved a sensitivity and coverage adequate for probing signatures of diffuse emission in the energy range of $\sim$100 GeV to a few TeV. $\gamma$-rays are produced in cosmic-ray interactions with the interstellar medium (aka ``sea of cosmic rays'') and in inverse Compton scattering on cosmic photon fields. This inevitably leads to guaranteed $\gamma$-ray emission related to the gas content along the line-of-sight. Further contributions relate to those $\gamma$-ray sources that fall below the current detection threshold and the aforementioned inverse Compton emission. Based on the H.E.S.S. GPS, we present the first observational assessment of diffuse TeV $\gamma$-ray emission. The observation is compared with corresponding flux predictions based on the HI (LAB data) and CO (as a tracer of H$_2$, NANTEN data) gas distributions. Consequences for unresolved source contributions and the anticipated level of inverse Compton emission are discussed. }


\begin{document}
\maketitle
\section{Introduction}
The Galactic diffuse $\gamma$-ray emission, observed by several space-born $\gamma$-ray experiments like SAS-2 \cite{SAS2}, COS-B \cite{COSB}, EGRET \cite{EGRET}, and Fermi-LAT \cite{FermiDiffuse}, is the most prominent feature of the sky at GeV energies. The processes leading to this emission are cosmic-ray interactions: bremsstrahlung and $\pi^0$ production in the interstellar medium and inverse Compton scattering on radiation fields. The contri\-butions differ in their relative amplitude on different energy scales and vary in their spatial extension depending on the interaction targets as the matter density and the respective radiation fields. When moving to higher energies, discrete $\gamma$-ray sources become more prominent compared to cosmic-ray induced processes (at TeV energies, inverse Compton and $\pi^0$ decay) and dominate the celestial $\gamma$-ray emission. The TeV energy regime is the realm of ground-based observations. Diffuse $\gamma$-ray emission has been reported by the Milagro experiment \cite{MilagroDiffuse} at a median energy of 15~TeV, and by ARGO-YBJ \cite{ARGO}. Common to both experiments is that while providing a very good duty cycle and a large field of view, they suffer from a rather poor angular resolution, which challenges the identification of discrete $\gamma$-ray sources. 
With a lower energy threshold and arc-minute angular resolution, imaging atmospheric Cherenkov telescopes have the potential to improve on these measurements at energies star\-ting at $\sim 100$~GeV and thereby improve our understanding of the Galactic diffuse $\gamma$-ray emis\-sion and its underlying production mechanisms. The most suited experiment for a measurement of diffuse emis\-sion in the Galactic Plane is the High Energy Stereoscopic System (H.E.S.S.): with its location in Namibia it provides ideal viewing condi\-tions on the central part of the Galactic Plane and features a comparatively large field of view for imaging atmospheric Cherenkov telescopes of $5^\circ$ in dia\-meter. 
The H.E.S.S. experiment has performed a survey of the Galactic Plane and has from 2004 to 2013 accumulated about 2800 observation hours of good-quality data. Although the H.E.S.S. Galactic Plane Survey (GPS) has revealed a wealth of new sources \cite{Catalog}, and already resolved dif\-fuse emis\-sion in the Galactic Center region on sub-degree \mbox{scales} \cite{Diffuse2006}, a measurement of the large-scale dif\-fuse $\gamma$-ray emission is at the edge of current instrumental sensitivity and a challenge for analysis methodology.
 
 \begin{figure*}[!t]
  \centering
  \includegraphics[width=\textwidth]{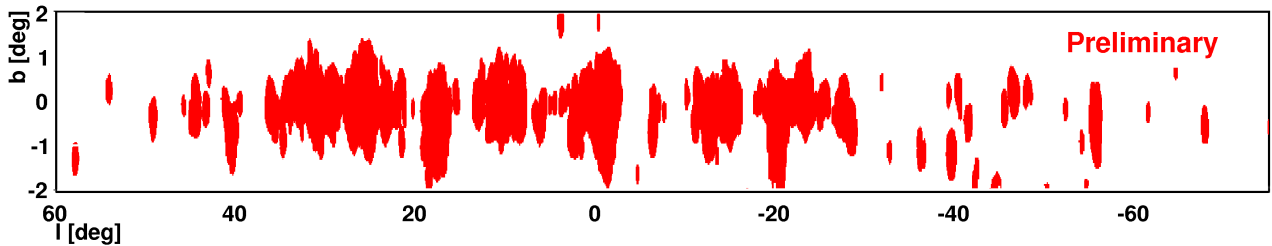}
  \caption{The region in Galactic longitude and latitude used for the measurement of diffuse $\gamma$-ray emission. Red denotes exclusion from the analysis because of the presence of significant $\gamma$-ray sources, white is in the following denoted ``diffuse analysis region''.}
  \label{fig1}
 \end{figure*}

\section{Methodology}
\subsection{Background subtraction}
A measurement of diffuse $\gamma$-ray emission is challenging for imaging atmospheric Cherenkov telescopes because of their restricted capabilities of $\gamma$-hadron separation and the limited field of view of a few degrees. The limitations in the $\gamma$-hadron separation require a {\it subtraction} of the background. In order to minimize systematic ef\-fects due to changing atmospheric or instrument conditions, the background level is determined from mea\-surement, usually within the field of view, in regions of no known $\gamma$-ray sources. This has proven to be the most reliable and fail-safe background-subtraction technique \cite{BGBerge} and is therefore used throughout the community. However, this method places severe constraints on the size of emission that can be probed. Furthermore, for emission larger than the field of view, the procedure results necessarily in a subtraction of part of the signal together with the background. Therefore, special care needs to be taken in the treat\-ment of the background and the interpretation of signals obtained after background subtraction.\\
For the current analysis a dataset of $1926$ hours of dead-time corrected obser\-vations are used, 
covering a region of $-75^\circ < l < 60^\circ$ in longitude and $-2^\circ < b < 2^\circ$ in la\-titude.
The data were analysed using the Model analysis technique \cite{Model} with standard cuts for event reconstruction and background reduction. The remaining background was measured in regions that do not meet any of the following exclusion criteria:
\begin{enumerate}
\item Any region is excluded that contains a $\gamma$-ray signal with significance of $>4\sigma$ in the chosen analysis bin and a significance in one neighbouring bin of $>4.5\sigma$ or lies within $0.2^\circ$ of such a signal (in order to include also tails in the point spread function used to describe $\gamma$-ray sources). Significances are consi\-dered for stan\-dard and hard cuts \cite{Model} and regions are determined iteratively. The resulting regions are visualized in Fig.~\ref{fig1}.
\item Galactic longitudes within a latitude range of $|b|<1.2^\circ$ are excluded in order to study a diffuse signal close to the Galactic Plane.
\end{enumerate} 
The first criterion is a very conservative approach assuring a minimum of $\gamma$-ray-source contaminations.
The choice of the latitude range in the second criterion is a compromise between a desired large excluded region in order to avoid a contamination of the background estimate on the one hand and the need for sta\-tis\-tics and reduction of systematics in the background measurement on the other hand. It is further motivated by the scale height in the distribution of interstellar gas, which is expected to correlate to a certain level with p-p interaction induced $\gamma$-ray emis\-sion. An adaptive ring background subtraction method has been chosen \cite{Catalog} for an optimal treatment of these constraints in the choice of background regions.\\
It is worth noting here that a subtraction of the background measured at regions with $|b|>1.2^\circ$ also means a subtraction of any large-scale diffuse signal that extends signifcantly beyond $|b|>1.2^\circ$. Therefore, any signal measured at $|b|<1.2^\circ$ is a signal with respect to the baseline of a potential flux outside the region.\\
After background subtraction the $\gamma$-ray excess events are folded with the exposure to obtain a 2-dimensional re\-pre\-sentation of flux in the Galactic Plane \cite{Catalog}.

\subsection{Discrete $\gamma$-ray sources}
As the flux in the Galactic Plane is completely dominated by many, mostly extended $\gamma$-ray sources, a measurement of diffuse $\gamma$-ray emission needs to exclude these sources from the analysis. Due to their spatial extension and the limitations in properly characterizing their flux distributions, a modelling of the sources turns out to be presently not feasible. Therefore, source locations are identified and excluded from the analysis by applying a cut in the observed detection significance. The same criterion (1) as for the choice of the background regions is applied. The remaining regions in the sky are de\-noted as ``dif\-fuse analysis region'' in the following. Results are stable with respect to the details in the choice of the analysis region.\\ 
The procedure of spatially excluding any significant signal results by definition in fluxes (as function of position) that are individually not significant. However, by investigating profiles of the flux distribution, the cummulative projected signal results in a notable flux excess.
\section{The diffuse $\gamma$-ray signal}
\begin{figure}[t]
  \centering
  \includegraphics[width=0.45\textwidth]{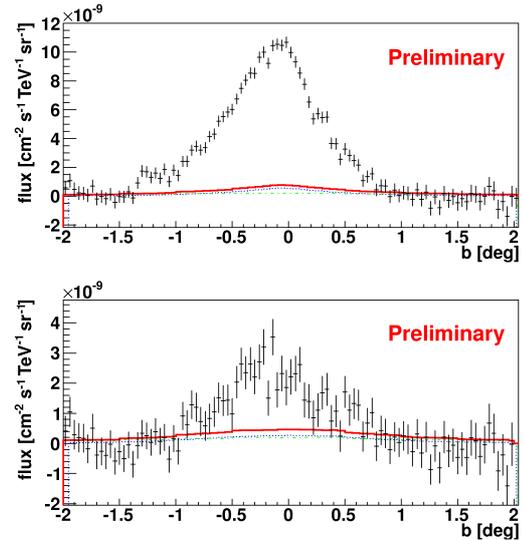} 
  \caption{The latitude profile of $\gamma$-ray flux (shown is the differential flux at an energy of 1~TeV), covering a longitude range of $-75^\circ<l<60^\circ$, for the total flux including $\gamma$-ray sources (top panel) and for the diffuse analysis region only as defined in Fig.~\ref{fig1} (bottom panel). H.E.S.S. measurements (black data points) are compared with the calculated level of $\gamma$-ray emission due to p-p interaction of the locally measured cosmic-ray spectrum with HI and H$_2$ (solid red line, the individual components are blue dotted for H$_2$ and green dot-dashed for HI interactions).}
  \label{fig2}
 \end{figure}

The latitude flux profile of the Galactic Plane for a longitude range of $-75^\circ<l<60^\circ$ is shown in Fig.~\ref{fig2} for both, fluxes including  $\gamma$-ray sources and fluxes of the diffuse analysis region only. Errors are $1 \sigma$ statistical errors and do not account for systematics.\\
Both distributions are characterized by a clear excess in the proximity of the Galactic Plane. Notebly, also the excess of the di\-ffuse analysis region is significant. It peaks not exactly at $b=0^\circ$ but slightly offset at around $b=-0.25^\circ$ with a peak value of around $2.5\times 10^{-9}$~cm$^{-2}$~s$^{-1}$~TeV$^{-1}$~sr$^{-1}$. The signal accumulates over the considered longitude range and consists of larger contributions from longitude values of the Galactic centre region and smaller ones from the outskirts of the observed region.\\\\
The observed diffuse emission can be interpreted as a combination of contributions from
\begin{itemize}
\item unresolved $\gamma$-ray sources
\item $\gamma$-rays resulting from cosmic-ray interactions with the interstellar medium ($\pi^0$ decay)
\item $\gamma$-rays resulting from cosmic-ray interactions with radiation fields (inverse Compton scattering).
\end{itemize}
The estimation of the individual contributions to the observed signal is non-trivial and further complicated by the relatively small signal.
A large part of the signal can be expected to stem from faint sources that are unresolved because of their low fluxes (avoiding significant detection) or very large extension not properly handled in the standard analysis.\\ 
A minimum level of cosmic-ray induced contribution can be estimated from p-p interactions with interstellar matter.
The $\gamma$-ray flux can be calculated following \cite{CRBarometers}
\begin{equation}
\frac{dN_\gamma}{dA dE_\gamma dt d\Omega} = \int d l_d \int \frac{d\sigma_{p\longrightarrow \gamma}}{dE_p} n(l,b,l_d) J(E_p)dE_p
\end{equation}
with $l_d$ being the line-of-sight, $E_p$ the proton energy, $J(E_p)$ the cosmic-ray spectrum, $\frac{d\sigma_{p\longrightarrow \gamma}}{dE_p}$ the interaction cross section for the $\gamma$-ray producing interaction, and  $n(l,b,l_d)$ being the column density of the gas of the interstellar medium. The cosmic-ray spectrum used for the calculation is the one measured locally at Earth, taken from \cite{PDG}. The interaction cross section is a parametrization of the SI\-BYLL interaction code following \cite{Kelner}. The interstellar matter that constitutes the target material are HI and  H$_2$. HI data are measurements from the Leiden/Argentine/Bonn survey (LAB \cite{LAB}) assuming a spin temperature of $T=125$~K, the H$_2$ column density is obtained using NANTEN CO data \cite{Nanten} as tracer of H$_2$. A conversion factor of $X_{CO} = 2\times10^{20}$~cm$^{-2}$~K$^{-1}$~km$^{-1}$~s \cite{XCO} has been used to convert the velocity-integrated NANTEN data to H$_2$ column density. The results of these calculations can be seen in the mo\-del curves of Fig.~\ref{fig2}. In order to assure comparability, the same regions in the sky have been used for the calculation of the expected $\gamma$-ray signal and the analysis, i.e. positions of $\gamma$-ray sources have been excluded from the model calculation for the diffuse analysis region as well. Note that due to the poor angular resolution of the HI data sources are excluded on scales smaller than the HI bin size, which is justified only by the apparent lack of correlation between $\gamma$-ray sources and HI densities.\\
The calculated $\gamma$-ray emission from p-p interactions has to be treated as a {\it minimal} or {\it guaranteed} level of the anticipated diffuse emission signal. The calculation uses the locally measured cosmic-ray spectrum, while the flux is expected to vary throughout the Galaxy. The flux observed at Earth, in no immediate proximity of any cosmic-ray accelerator is assumed to be the minimum level of cosmic rays in the Galaxy (``sea of cosmic rays''), while close to accelerators the level can be significantly enhanced.\\ 
Further simplifying assumptions that are invoked (most of them reducing the estimated contribution) include a constant $X_{CO}$ (as opposed to some functional dependence on Galactocentric distance), the limitation to hydrogen contributions and usage of p-p cross section also for heavier cosmic rays.\\
As the contribution of neutral-pion-decay $\gamma$-rays is neatly localized along the Galactic Plane, a comparison of the contribution can be made with the H.E.S.S. data in a first approximation without considering the issue of background subtraction and the corresponding reduction of the signal. In the region of $|b| < 1^\circ$ in the diffuse analysis region the integrated contribution of the calculated $\gamma$-ray signal originating from $\pi^0$ decay is $\sim$25\% of the H.E.S.S. measurement (as seen in Fig.~\ref{fig2} bottom panel), thereby limiting the contribution of unresolved sources (also expected to concentrate close to the Plane) to less than 75\%. In comparison, the calculated contribution from neutral-pion-decay $\gamma$-rays to the total flux including $\gamma$-ray sources is for the region $|b| < 1^\circ$ less than $10\%$ (as in Fig.~\ref{fig2} top panel). \\
The second cosmic-ray interaction component is inverse Compton scattering. Cosmic-ray electrons upscatter photons of optical starlight, infrared dust emission and the cosmic microwave background. The cosmic-ray electrons that are responsible for the inverse Compton emission at TeV energies have, depending on the radiation field they are interacting with, energies between a few and a few hundred TeV. At these energies, energy losses are severe and their lifetimes and propagation distances are limited. Therefore, the electrons are found close to their production sites, which are either homogeneously distributed in case of secondary production of electrons via hadronic cosmic-ray interactions, or highly inhomogeneously localized as for primary cosmic-ray electrons \cite{Pohl}. This makes an estimation of a contribution to the diffuse $\gamma$-ray emission at TeV energies challenging and dependent on source model assumptions.\\
Since the radiation fields as target for inverse Compton  scatter\-ing extend to higher latitudes than the bulk of the gas distribution, the inverse Compton component to be contained in the diffuse $\gamma$-ray signal will be subject to background subtraction and only gradients will be measurable in the observed signal.

\section{Conclusion}
We present the first ever investigation of large-scale diffuse emission with imaging atmospheric Cherenkov telescopes. Imaging atmospheric Cherenkov telescopes have the advantage of a rather precise direction reconstruction, which puts them into advantage compared to other instruments that measure at TeV energies in the identification and exclusion of $\gamma$-ray sources as dominant ``background'' of a diffuse $\gamma$-ray analysis. The H.E.S.S. data reveal an excess around a latitude of about $b=-0.25^\circ$ over the baseline at latitude extensions larger $1.2^\circ$ in the Galactic Plane. This signal exceeds the guaranteed contribution stemming from the calculated level of $\gamma$-ray emission due to p-p interaction of the locally measured cosmic-ray spectrum with HI and H$_2$ via $\pi^0$ decay. Additional contributions to the observed signal relate to unresolved $\gamma$-ray sources and inverse Compton sca\-ttering of cosmic-ray electrons on interstellar radiation fields. \\\\

{\small The support of the Namibian authorities and of the University of Namibia
in facilitating the construction and operation of H.E.S.S. is gratefully
acknowledged, as is the support by the German Ministry for Education and
Research (BMBF), the Max Planck Society, the German Research Foundation (DFG), 
the French Ministry for Research,
the CNRS-IN2P3 and the Astroparticle Interdisciplinary Programme of the
CNRS, the U.K. Science and Technology Facilities Council (STFC),
the IPNP of the Charles University, the Czech Science Foundation, the Polish 
Ministry of Science and  Higher Education, the South African Department of
Science and Technology and National Research Foundation, and by the
University of Namibia. We appreciate the excellent work of the technical
support staff in Berlin, Durham, Hamburg, Heidelberg, Palaiseau, Paris,
Saclay, and in Namibia in the construction and operation of the
equipment.}


\end{document}